\documentclass[prb,twocolumn,showpacs]{revtex4}\usepackage{graphicx}
\usepackage{color}
\begin{document}
\title{ Formation of new phase inclusions  in the system of quasi-equilibrium  magnons of high density}
\author{V.I.~Sugakov}
\affiliation{Institute for Nuclear Research, National Academy
of Sciences of Ukraine (47, Nauki Ave., Kyiv 03680,
Ukraine)}\email{sugakov@kinr.kiev.ua}
\begin{abstract}
The paper studies the spatial variation of the magnetization in a
nonconducting magnetic sample with an excess number of magnons in
comparison to the equilibrium. The phenomenon is considered using
the Landau-Lifshits equation with additional terms describing
 the longitudinal relaxation
of the magnetization, the magnon diffusion and the magnon creation
by external pumping. The free energy of the system is presented in
the mean field approximation. It is shown that, if the pumping
exceeds some critical value, regions of the new phase arise  where
the magnetic moments are oriented opposite to the magnetization of
the magnetic sample. The phenomenon is similar to the appearance
of droplets of the condensed phase in the supersaturated vapor.
The appearance of the new phase either in the form of a single
domain or a periodical lattice is demonstrated. The studied
process is a competitor to the process of the Bose-Einstein
condensation of magnons.
\end{abstract}

\maketitle

\section{Introduction}
Spatial structures in non-equilibrium nonlinear systems have been studied widely and successfully during the last half century\cite{N}.
 The processes of the formation of such structures are referred to as the non-equilibrium phase transitions.
 Non-equilibrium phase transitions are various
and in their majority differ from the equilibrium phase
transitions. There is a class of specific non-equilibrium phase
transitions in systems of particles (or quasiparticles in
crystals), which have finite values of  the lifetime. If combining
particles produces an energy gain  and the lifetime of particles
is much larger than the time of inter-particle collisions, the
particles may form a condensed phase. The formation of drops of
the electron-hole liquid in germanium or silicon with high density
of excitons created by light is a classical example of the phase
transitions in  systems of unstable particles\cite{Keld,Rice}. The
stationary state of the condensed phase of unstable particles may
exist  only in the presence of a source which creates new
particles instead of the disappearing ones. When the value of the
lifetime is large, the parameters of the condensed phase (density,
critical temperature of the phase transition etc.) are only
slightly modified compared to the same parameters for the infinite
value of the lifetime.  But for the range of parameters, where the
gas and the condensed phase coexist, the spatial distribution of
the finite lifetime particles has particular features which will
be discussed later.

Magnons in magnetic materials are the classical example of
particles with the finite lifetime. The chemical potential of
magnons is equal to zero in the equilibrium state because their
number is not conserved due to the magnon-magnon and magnon-phonon
interactions. In presence of an external pumping, additional
magnons appear besides the equilibrium magnons. As the result, the
chemical potential is not equal to zero. Magnons are Bose
particles and when their concentration exceeds some threshold
value, the appearance of the Bose-Einstein condensation (BEC)
could be expected together with its interesting manifestations:
accumulation of particles at a certain level, superfluidity and so
on. An interesting effect was observed in the
works\cite{Dem1,Dem2} in which the authors investigated magnons in
the yttrium-iron-garnet (YIG) films. The magnons were excited by
the parametric longitudinal  pumping. The analysis of the magnon
spectra was carried out using Brillouin light scattering (BLS)
spectroscopy. The authors showed from the analysis of experiments
the manifestation of the Bose-Einstein condensation of magnons. In
particular, an increase of the magnon concentration in the  state
with the wave vector which corresponded to the minimum of the
magnon band was observed\cite{Dem1,Dem2}. The appearance of the
spontaneous coherence in BLS by magnons was observed if the
pumping exceeded a critical value \cite{Dem3}. The emergence of
the periodical variation of the magnon density at a high level of
the magnon excitation was also demonstrated \cite{Dem4}. The
latter phenomenon was explained by the presence of two minima in
the magnon dispersion law leading to the formation of two
condensates and to the interaction between the condensates.

Since then the investigation of the magnon condensation in YIG
 obtained further development in numerous works
 which further advanced
the explanation of the phenomenon observed in
\cite{Dem1,Dem2,Dem3,Dem4}  and suggested different other effects
related to Bose-Einstein condensation. Thus, the stability of BEC
in the high density magnon system in YIG was analyzed in the paper
\cite{Tup}. The microwave emission from the uniform mode generated
by BEC studied in \cite{Rez}. The spatial structure of interacting
bosons with two minima in the dispersion law was investigated in
the papers \cite{Hick,Lok2,Li}. The dramatic peak in the density
of the proposed condensed magnons after switching off the pumping
was observed in the work \cite{Serga}. It is interesting that the
time by the which the peak increased coincided  with the time of
the magnon decay and the time by the peak decreased was much
larger than the magnon decay time.
 The problem of the spin
current in the system  of an isolator and a conductor was
theoretically studied  under the condition of BEC of magnons in
the isolator in \cite{Duin}. In the papers \cite{Nak,Tron} the
Josephson oscillations in the magnon density between two spatially
separated magnon clouds were calculated and also the methods of
the magnon current measurement are analyzed. In the paper
\cite{Bozh}  the temporal decrease of the magnon condensate
density in YIG at the gradient of the temperature, created by a
laser, after the pumping shutdown was observed. The authors
explained this effect by the appearance of a supercurrent in the
condition of BEC.
 But there are  works in which the doubt is expressed about
the correctness of the interpretation of the results observed in
\cite{Dem1,Dem2}. The authors of the work \cite{Lok1} showed that
the reason of an accumulation of  particles created by pumping at
the lowest state may be caused by the peculiarities
 of the Bose-Einstein condensation of quasi-particles.
In the paper \cite{Ruck} the authors described the time evolution
of the magnon condensate under pumping by the classical stochastic
Landau-Lifshits-Gilbert equation including magnon-phonon
hybridization and came to  the conclusion that the phenomenon
observed in \cite{Dem1,Dem2} has a purely classical nature.

In the current paper, we present another version of the processes in
 a ferromagnet with the magnon
density exceeding the equilibrium value. We
 show that, if  there are
additional magnons, created by an external pumping,  the evolution
of the system may choose the scenario alternative to the
Bose-Einstein condensation. This new scenario is the formation of
regions  in the ferromagnetic material where the magnetic moments
are  oriented opposite to the orientation  of the magnetic moment
of the sample. Similar to the Bose-Einstein condensation, this
phenomenon appears in crystals with the magnon density higher than
the equilibrium value.

\begin{figure}\centerline{\includegraphics[width=8.6cm]{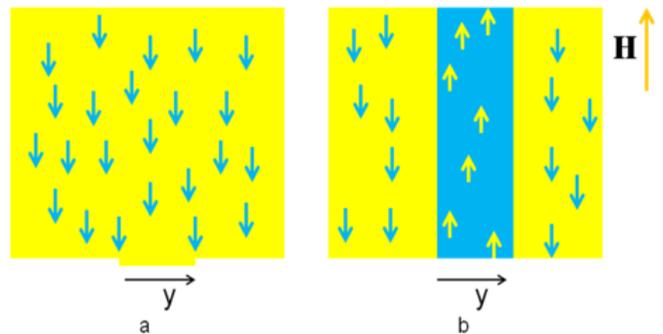}}\caption{(Color online) The
distribution of magnon magnetic moments in a crystal. The
continuous regions present the areas with majority of lattice cell
moments oriented along the field (light) and opposite to the field
(dark). The individual lattice cell moments are not shown because
there are too many of them. The arrows show the directions of the
moments of magnons which are opposite to the direction of the
majority of cell moments. The left part (a) presents the uniform
distribution of magnons at which the Bose-Einstein condensation
may occur if the magnon density is higher than some threshold
value. The right part (b) shows an alternative distribution of the
magnetization, in which the state with a domain (the dark strip)
having moments of the lattice cells oriented opposite to the
magnetic field arises. The light arrows in the dark region show
the magnetic moments of the magnons in the domain.}
\label{fig:fig1}\end{figure}

The system of magnons is in a way equivalent to a gas of
particles. If the concentration of the particles exceeds the
equilibrium value, the gas is referred to as "over-saturated" or
"supersaturated". The processes of the precipitation of the
regions of the new phase are known to occur in the over-saturated
gas. Similar systems arise also in  mixtures of liquids or solids
after rapid cooling. There are two popular models that describe
processes of the unmixing of mixtures from one thermodynamic phase
to form two coexisting phases: the model of the spinodal
decomposition\cite{Cahn,Hil} and the model of the nucleation and
growth\cite{Lif}. The subject of our interest, the system with the
magnon pumping, is "over-saturated" with magnons. So, during the
relaxation the individual magnons would  cluster forming
inclusions of the new phase. Within these inclusions, the
orientation of the magnetic moments would be opposite to the
magnetic moment of the crystal. A qualitative picture for the
dynamics of the magnon system is shown in Fig.~\ref{fig:fig1}.

 Let us assume that the magnon  state of  Fig.~\ref{fig:fig1}a presents the uniform quasi-equilibrium magnon distribution
  which arises due to both the thermal excitation and the external pumping. There are two scenarios for the further development
  of the uniform magnon distribution.
According to the first scenario, the Bose-Einstein condensation
shall occur if the magnon concentration  exceeds the critical
value. Such process is investigated in the cited
papers\cite{Dem1,Dem2}. But, the second scenario, according to
which the formation of the new phase occurs in the over-saturated
magnon system, is also realistic. There is a strong short-range
interaction between magnons. When magnons are collected in a
cluster, the energy per magnon decreases by the value of order of
0.1~eV \cite{Gur}, which significantly exceeds  the thermal energy
at the room temperature. The equilibrium state of the system is
determined by the minimum of the free energy and not by the
minimum of the energy, But, if the  magnon concentration exceeds
the equilibrium one, the magnon clusters have to form with the
opposite orientation of their magnetic moments to the magnetic
moment of the other part of the crystal. The magnon clusters are
inclusions of regions of the magnon condensed phase. But it is not
the Bose-Einstein condensation, it is the conventional
condensation in the coordinate space due to the interaction
between particles. The clusters of the new condensed phase may be
shaped variously. A structure in the form of a single domain is
drawn in Fig.~\ref{fig:fig1}b. The presented paper investigates
the processes of the formation of the regions of the new phase in
the supersaturated magnon gas.

Because the magnon lifetime is finite, the arising structures may
exist  only during the continuous magnon pumping. Similar studies
of the phase transitions in the systems of particles with the
finite lifetimes have been carried out for different types of
quasi-particles: for the radiation defects\cite{Sug1,Sel, Russ},
for the excitons\cite{Sug2,Sug3,Ogawa,Ish,Chern1,Sug4}. The
theories of these works have modified and generalized the
stochastic model of the nucleation and growth (Lifshiz-Slyosov
\cite{Lif}) and the model of the spinodal decomposition
(Cahn-Hillert \cite{Cahn,Hil}) to make them applicable to systems
of particles with a finite lifetime. The theories have been
successful in explaining the unconventional experimental results
obtained by different authors (mainly by the Timofeev's\cite{Tim}
and Butov's\cite{But} groups) during investigation of the light
emission by excitons from the double quantum well heterostructures
in semiconductors at low temperature. Excitons were created by
lasers and, at high density, formed
 islands of the excitonic condensed phase. Sometimes the islands were localized periodically in the space.
  The references to the applications
of the theory of phase transitions in systems of unstable
particles to the explanation of experiments with excitons are
given in \cite{Chern1,Sug4,Chern}. The formation of the new phase
in a system of unstable particles has distinct features compared
to the phase transition in the system of stable particles. The
distinctions include: 1) the size of the regions of the new phase
is restricted, 2) there is a correlation between the regions of
the new phase, which may cause the appearance of periodical
structures, 3) the regions of the new phase exist only at the
external pumping. The structures are the results of
self-organization processes in non-equilibrium systems.

In the presented paper, we apply to the many-magnon system the
approach developed in the above  papers (the paper \cite{Chern}
and the references therein) devoted to the study of the phase
transitions in systems of unstable particles. We shall show the
possibility of an appearance of the new phase inclusion  in the
magnetic with supersaturated magnon gas. The qualitative analysis,
given in the last section, argues  that effects caused by the new
inclusions may be similar to the effects, which were observed in
\cite{Dem1,Dem2,Dem3,Dem4} and explained by the manifestation of
BEC.

\section{Taking into account magnon diffusion and magnon pumping into equation for magnetization  }

We shall consider a nonconducting magnetic crystal in the magnetic
field oriented along the crystalline axis $0Z$. We assume that the
non-equilibrium magnons are excited in the system by the
two-magnon longitudinal pumping and the number of magnons is
larger than the equilibrium one. Due to the strong magnon-magnon
and magnon-phonon interactions, the magnons are in the
quasi-equilibrium state.   Our aim consists in the determination
of the spatial variation of the magnetization $\mathbf{M}$. To
this end, we shall study  the clustering of magnons into a new
phase with the creation of regions that have the magnetic moment
oriented opposite to the orientation of the main magnetization.
 After the creation of the regions of the
new inverse phase, the magnetization is non-uniform, though the
initial system and external fields (the static magnetic field, the
pumping) are assumed to be uniform. The processes of the
self-organization in the system spontaneously break the symmetry.
Describing the non-uniform system we assume that the principle of
the local equilibrium holds. In this case, in a small vicinity of
some spatial point, the thermodynamic functions are the same
functions of the local microscopic variables (magnon density,
temperature) as in the equilibrium system. This assumption allows
introduction of the free energy in non-equilibrium system. The
local free energy depends on the magnon density and the magnon
density depends on the spatial coordinates. The principle of the
local equilibrium is conventionally used in the majority cases
when considering the  self-organization problems in
non-equilibrium systems \cite{N}.

We shall use the phenomenological
approach for the solution of the problem. Let us analyze the phenomenological equation for the magnetization $\mathbf{M}$, solution of which  will be
investigated in the paper.  The equation for the change of the
magnetization in the unit time contains the dynamic and the relaxation
parts. The
 relaxation
terms of the Landau-Lifshits (LL) and the Landau-Lifshits-Gilbert
(LLG) equations cannot be used in our paper because they require
the conservation of the magnetization. The pumping creates magnons
and decreases  the absolute value of the magnetization. The LL and
LLG equations do not describe the equilibration of the
magnetization after the pumping is switched off. Taking into
account  the processes of the establishment of the equilibrium
state  is important in a description  of the non-equilibrium
system. The equation for the evolution of the magnetization which
do not  require the conservation of the absolute value of the
magnetization is given in the monograph of Akhiezer, Baryakhtar
and Peletminskii \cite{Akhiezer}. But neither the equation in
\cite{Akhiezer} and nor the LL and LLG equations take into account
the diffusion of magnons. The diffusion processes induce spatial
redistribution of magnons due to the interaction between them,
which may be a reason for the formation of the new phase. The
diffusion is important in the formation of the new phase at
spinodal decomposition processes \cite{Lif,Cahn,Hil}.
 So, we presented the main equation for
the magnetization in the form given in \cite{Akhiezer} and added
to its right hand side the terms describing the magnon diffusion
$\left(
\partial \mathbf{M}/{\partial t}\right)_D$ and the pumping
$\mathbf{P}$.
 \begin{equation}
\label{eq1} \left (\frac{\partial \mathbf{M}}{\partial
t}\right)=-\gamma
[\mathbf{M},\mathbf{H_{eff}}]+\mathbf{R}+\mathbf{P}
\end{equation}
where $\gamma$  is the gyromagnetic ratio, $\mathbf{H_{eff}}$ is
the effective magnetic field determined as the variational
derivative of the free energy with respect to the magnetization
\begin{equation}
\label{eq2}
 \mathbf{H_{eff}}=-\frac{\delta F}{\delta \mathbf{M}},
\end{equation}
$\mathbf{R}$ is the relaxation term
\begin{equation}
\label{eq2a}
 \mathbf{R}=-\gamma_{R1}[\mathbf{n_M},[\mathbf{n_M},\mathbf{H_{eff}}]]
+\gamma_{R}\mathbf{H_{eff}}+\left (\frac{\partial
\mathbf{M}}{\partial t}\right)_D,
\end{equation}
 $\gamma_R$ and $\gamma_{R1}$ are the relaxation rates,
$\mathbf{n}_M=\mathbf{M}/M$.

The two first terms in the right hand side of Eq.~(\ref{eq2a})
determine the relaxation in the book \cite{Akhiezer}.
 The first term in  Eq.~(\ref{eq2a}) is equal to the relaxation term
 of the LL equation. For a such relaxation, the conservation of the absolute value of the magnetization holds.
  The LLG equation for the magnetization  is equivalent to the LL
  equation:  the LLG equation may be obtained from the LL
  equation by redefining  parameters \cite{Bert}. Therefore, the first term
 cannot describe the magnetization  in the condition of the external magnon pumping. The second and the third terms in  Eq.~(\ref{eq2a})
 are important for the description of the development of  the non-uniform
 variation of the magnetization under pumping. The third term in Eq.~(\ref{eq2a}), responsible for the
 magnon diffusion, is absent from the LL, LLG and Akhiezer, Baryakhtar
and Peletminskii \cite{Akhiezer} equations.

To describe the contribution of the exchange interaction into the
free energy, we use the method of the self-consistent field and
present the free energy in the form
 \begin{eqnarray}
\label{eq3}
F(\mathbf{M},\mathbf{\nabla}\mathbf{M})=\frac{aM^2}{2}+\frac{bM^4}{4}+\frac{K}{2}(\mathbf{\nabla}\mathbf{M})^2
\nonumber\\-MH\cos\theta
-\frac{1}{2}{\mathbf M} \cdot {\mathbf H}^{(m)}+K_{an}\sin^2\theta ,
\end{eqnarray}
where $K_{an}$ is the anisotropy constant, $\theta$  is the angle
between the magnetic field and the crystal axis, ${\mathbf H}^{(m)}$
is the magnetic field created by the magnetic moment ${\mathbf M}$,
$a$, $b$ and $K$ are parameters depending on the temperature and
not depending on the spatial coordinates. The first three terms
describe the exchange interaction. The free energy  may be given
in the form of Eq.~(\ref{eq3}) at the temperature close to the
critical temperature of the phase transition. Such case will be
studied in this paper.

As mentioned, the regions of the new phase, appearing due to the pumping,
may  have different shapes. We shall consider the formation of a
magnetic domain with the orientation of the magnetic moments
opposite to the orientation of the magnetic moment of the crystal
and parallel to its easy-axis. Domains of such type are simple and
widespread defects in magnetics.

 In the transition area, where
one orientation of the magnetic moments change to another, the
exchange energy increases. The transition may occur either by a rotation
of the magnetic moment which preserves its absolute value or by
changing the value of the moment. Usually the first way dominates
due to the high magnitude of the exchange interaction. But in the
vicinity to the phase transition, the exchange interaction
decreases and the second way is plausible
 (see the textbook problem in \cite{Landau}). Since we
study the processes nearby the critical temperature, the
transition between the domain and the matrix is considered by
changing the value of the magnetic moment without its rotation.

In the framework of the chosen approach, when the normal to the
domain plane is perpendicular to the magnetic field $\mathbf{H}$, the
orientation of the magnetic moments has the form presented
schematically in Fig.~\ref{fig:fig1}. The magnetic moment inside
the domain has the single component $M_z$ and depends on the
single variable $y$. Therefore, $\mathbf{M}\longrightarrow M_z(y)$
and $\mathbf{H}^{(m)}=0$. For a such orientation of the magnetic moment,
the angle $\theta$ in Eq.~(\ref{eq3}) is zero  and both the first term
in the right hand side of Eq.~(\ref{eq1}) and the first term in the right hand side of Eq.~(\ref{eq2a}) disappear.

Let us apply the phenomenological approach to determine how the
diffusion contributes to the time evolution of the magnetic
moment. The density of the magnon current in the non-homogeneous
system at the uniform distribution of the temperature may be
expressed by the gradient of the chemical potential $\mu$
   \begin{equation}
\label{eq4}
 \mathbf{j}=-K_M\mathbf{\nabla}\mu,
\end{equation}
where the coefficient  $K_M$ (mobility) depends on the
temperature. $K_M$ may be evaluated from the Boltzmann equation  for
the magnon distribution function taking into account the magnon
scattering on magnons, phonons and impurities.

Let us introduce the tensor of magnetization current $\Pi_{Mik}$ ,
that describes the density current of the $i$-th component of the
magnetic moment when the  magnons are moving along the axis $k$.
It is equal to the product of the $i$-th component  of the
magnetic moment of a single magnon and  $k$-th component of the
magnon current density
   \begin{equation}
\label{eq5}
 \Pi_{Mik}=m_ij_k.
 \end{equation}
The single magnon has the following magnetic moment
   \begin{equation}
\label{eq6}
 \mathbf{m}=-g\mu_B \mathbf{n_M}.
 \end{equation}
The vector $\mathbf{n_M}$  is oriented along the axis $z$. It may
assume two values $\pm 1$ . The sign "+" takes place in the matrix
where the moment is oriented along the magnetic field. The sign
"-" is realized in the domain. The rate of the change of the
magnetization is equal to
\begin{equation}
\label{eq7}
  \left( \frac{\partial \mathbf{M}}{\partial
t}\right)_{Di}=-\frac{\partial \Pi_{Mik}}{\partial x_k}.
 \end{equation}

 The magnon current creates the magnetic moment current
  \begin{equation}
\label{eq8}
  \Pi_{Mik}=-g\mu_B n_{Mi}j_k .
 \end{equation}
In the considered case, the tensor of the magnetization current has a single
nonzero component $\Pi_{Mzy}$. The contribution of the magnetic
moment current into the rate of the magnetization change is
equal to
 \begin{equation}
\label{eq10}
  \left (\frac{\partial M_z}{\partial
t}\right)_{Dy}=-\frac{\partial \Pi_{Mzy}}{\partial
y}=g\mu_B\frac{\partial }{\partial y}\left(n_{Mz}K_M\frac{
\partial\mu}{\partial y}\right).
 \end{equation}
The chemical potential may be determined via the free energy
by the formula $\mu= \delta F/ \delta n$ at the constant
temperature, where $n$ is the magnon density. The magnon density
may be related to the magnetization by the expression
  \begin{equation}
\label{eq11}
 \mathbf{M}=\mathbf{n_M}(M_s-g\mu_B n),
 \end{equation}
where the saturation magnetization $M_s=N\mu_B$, $N$  is the
number of the crystal cells in a unit volume. Eq.~(\ref{eq11})
does not take into account Walker's modes. But since we
consider the  high temperature case, the main contribution into
the decrease of the magnetization comes from the magnons with the
high value of the wave vectors. The authors  of the paper \cite{Vaks}
 showed that notion of the magnons as quasiparticles may hold
almost up to the temperature of the phase transition.

 Using the free energy of Eq.~(\ref{eq2})    and  Eqs.~(\ref{eq3}) and (\ref{eq11}) we obtain the chemical potential
 \begin{eqnarray}
\label{eq12} \mu=\frac{\delta F}{\delta \mathbf{M}} \frac{\partial
\mathbf{M}}{
\partial n}=- \mathbf{H_{eff}}(-g\mu_B\mathbf{n_M}) =
\nonumber\\-g\mu_B(aM+bM^3-n_{Mz}H-
K\mathbf{n_M}\triangle\mathbf{M}).
\end{eqnarray}

As seen from Eq.~(\ref{eq12}), the chemical potential may
be presented as the interaction energy of the effective magnetic
field   with the magnetic moment of a magnon
$(-m_Bg\mathbf{n_M})$. Without the pumping ($\mathbf{P}=0$),
Eq.(\ref{eq1}) has the solution $\mathbf{H_{eff}}=0$ and,
therefore,   $\mu=0$, as  follows from Eq.~(\ref{eq12}).

Let us consider now the contribution of the pumping into
Eq.~(\ref{eq1}). The rate of the magnon injection into the
sample depends on the method of the injection and numerous other conditions:
the frequency and amplitude of the external field, the state of
the magnetization of the sample, the magnon spectrum and so on
\cite{Mon,Gur}. We shall not consider the processes of the magnon
creation by the external microwave field. We assume only that  the
value of $P$ , which determines the change of the magnetization in
the unit time and the unit volume due to the pumping,  does not depend
on time or spatial position. The magnons created  by the pumping
come  very quickly to the quasi-equilibrium state due  to
the magnon-magnon and the magnon-phonon interactions and the dynamics of
system is determined by the total number of magnons. It is the major assumption of  the BEC studies as well. The pumping causes the
decrease of the magnetization. So, the value of $P$ should be
negative with respect to the crystal magnetization and positive in
the domain, where the magnetization changes sign, and should be
equal to zero at the point where $M_z=0$. We shall present the pumping
in a simplified form. Its value will be  approximated by the
formula
  \begin{equation}
\label{eq13}
 P=-qM_z.
 \end{equation}
where  $q$ is a positive phenomenological parameter. Its value is
deduced from the experimental decrease of the magnetization  due
to the pumping.

Using Eqs.~(\ref{eq10}), (\ref{eq12}) and (\ref{eq13}) in
Eq.~(\ref{eq1}) and taking into account that the dynamic term in
the considered case is  zero,  we obtain the equation for the
magnetization
\begin{equation}
\label{eq14} \frac{\partial M_z}{\partial t} = R_z+P.
\end{equation}
where
\begin{equation}
\label{eq15} Rz=\gamma_R H_{eff}- g\mu_BK_M \frac{\partial^2
H_{eff}}{\partial y^2},
\end{equation}
\begin{equation}
\label{eq16} H_{eff}=H-aM_z-bM_z^3+K \frac{\partial^2
M_z}{\partial y^2},
\end{equation}

Since the dynamic term in the Eq.~(\ref{eq14}) for  the magnetization
 is zero in the considered system, its right hand side consists of the relaxation term given by Eq.~(\ref{eq15}) and the
pumping. It is seen that dissipative term in Eq.~(\ref{eq15}) contains
the component with the second derivative of the effective magnetic
field. The form of the dissipative term $R_z$ coincides with the form
of the dissipative term $R_{Bar}$ obtained for the magnetization
in the Landau-Lifshits-Baryakhtar equation\cite{Bar1,Bar2,Bar3}
\begin{eqnarray}
\label{eq17} \left (\frac{\partial \mathbf{M}}{\partial
t}\right)=-\gamma [\mathbf{M},\mathbf{H_{eff}}]+\mathbf{R_{Bar}},
\end{eqnarray}
where
\begin{equation}
\label{eq18}
 \mathbf{R_{Bar}}=\lambda_1 \mathbf{H}_{eff}-\lambda_2\triangle\mathbf{H}_{eff}
\end{equation}
where $\lambda_1$ and $\lambda_2$ are tensor coefficients in
the general case.

Comparing the relaxation term
of Eq.~(\ref{eq15}) with the Baryakhtar's expression of Eq.~(\ref{eq18}),  we
obtain the values of the coefficients: $\lambda_1=\gamma_R$,
$\lambda_2=g\mu_BK_M$ for our particular case of the magnon diffusion.

Baryakhtar built the equation for the magnetization using the
Onsager kinetic equations and the crystal symmetry. As shown in a number of
works,  the terms additional to the LL and the LLG equations
are important for the explanation of many processes in
ferromagnets: the dynamics of domain walls\cite{Bar1,Wang}, the dynamics
of solitons\cite{Bar3}, the damping of the spin waves with the high values of the wave
vectors\cite{Bar2,Wang}, the anisotropic damping in the feromagnetics\cite{Dvor}, the temporal
 spin evolution in the magnetic heterostructures disturbed  by  femtosecond laser pulses\cite{Yas}. The general
 theory \cite{Bar1}
does not specify the numerical value of the coefficients. Their
values are determined for particular systems. For example, the
inclusion of the conductivity electrons in the magnetization
dynamics of conducting ferromagnets\cite{Wang,Zh} give the
additional terms that depend on the conductivity. The value of the
coefficient we obtained is determined by the diffusion of magnons.

Let us introduce the dimensionless variables
$\tilde{y}=y/l_0$,  $l_0=(K/(-a))^{1/2}$, $\tilde{M}=M/M_{00}$,
$\tilde{H}=H/H_{00}$, \qquad $\tilde{t}=Kt/(a^2g\mu_BK_M)$,
$\tilde{q}=qK/(a^2g\mu_BK_M)$, $\tilde{\gamma_R}=\gamma_R
K/((-a)g^2\mu_B^2K_M)$, where $M_{00}=((-a)/b)^{1/2}$ and
$H_{00}=(-a)(-a/b)^{1/2}$ are the value of the magnetization and the
effective magnetic field, respectively, in the absence of the
magnetic field and pumping.

Eq.~(\ref{eq14})  for the magnetization in the dimensionless
variables takes the form (the diacritic "$\sim $" above the
notations for $M_z$, $H$, $q$, $t$ and $y$ will be omitted from
now on)

\begin{eqnarray}
\label{eq19} \frac{\partial M_z}{\partial t}= \frac{\partial^2
}{\partial y^2}\left (-M_z+M_z^3-H- \frac{\partial^2 M_z}{\partial
y^2}\right)
 \nonumber\\-\tilde{\gamma}_R\left (-M_z+M_z^3-H- \frac{\partial^2 M_z}{\partial
y^2}\right)-q M_z.
\end{eqnarray}

Eq.~(\ref{eq19}) determines the variation of the magnetization in
the region of the phase transition in the presence of the magnon
pumping.  The first and the second terms in its right hand side
describe dissipative processes.   The first term originates from
the processes of the magnon diffusion.  The second term describes
the relaxation of the magnetization to the equilibrium value.

 Let us consider expressions which may be used for the estimation of the numerical values of parameters.
  The mobility $K_M$ may be related to
the magnon diffusion coefficient  by the Einstein's relation
$K_M=Dn/\kappa T$, where $T$  is the temperature. The diffusion
coefficient may be obtained from the solution of the Boltzmann
equation for magnons. According to Eq.~(\ref{eq11}),
$n=(M_s-M)/g\mu_B $. In the vicinity of the phase transition,
where $M \ll M_s$, we have $n\sim M_s/g\mu_B$. In the mean field
approximation, the parameters of the free energy \cite{Kop} are
\begin{equation}
\label{eq20} a=-\Delta t\frac{\kappa T_c}{M_s\mu_B}, b\sim
\frac{1}{3} \frac{\kappa T_c}{M_s^3\mu_B}, K\sim \frac{\kappa
T_c}{M_s\mu_B}d^2,
\end{equation}
where $T_c$  is the phase transition temperature, $\Delta
t=(T_c-T)/T_c$, $d$ is the period of the crystal lattice. In the
approach of Eq.~(\ref{eq20}) in dimensionless variables,  the
length unit is equal to $l_o=d/(\Delta t)^{1/2}$,  the
magnetization unit is  $M_{00}=(\Delta t)^{1/2}\sqrt{3}M_s$, the
magnetic field unit is equal to $H_{00}=M_s(\Delta
t)^{3/2}\sqrt{3}(\kappa T_c)/(\mu_BM_s)$. Let us do the
estimations of  the values of the parameters  in the dimensionless
units. We consider the parameters of  the yttrium-iron-garnet
crystal, in which $T_c=560$~K, $M_s=140$~Gs. We assume that
$D=100$~cm$^2$/s. Since the dimensionless magnetization is the
ratio of the magnetization to the value of the magnetization of
the sample without the magnetic field and pumping,   its magnitude
is of order of unity or smaller ($\tilde{M}\leq 1$). The magnetic
fields which varies in the experiments \cite{Dem1,Dem2,Dem3} from
0 and to $1000$~Oe is described in the dimensionless units by the
values that are less than the unity  ($\tilde{H}\ll 1$). For
example, the magnetic field of 1000~Oe is equal in the
dimensionless units to 0.00077 and 0.0022 at the temperatures
$\Delta t=0.2$ and $\Delta t=0.1$, correspondingly. We shall
choose the parameter of the pumping  $q$ in Eq.~(\ref{eq13}) in
the way that ensures a small decrease of the magnetization due to
the pumping  (less than several percent of  the  magnetization
value). The dimensionless coefficient $\tilde{\gamma}_R$ is found
to be very small. For $\Delta t=0.1$, varying the parameter
$\gamma_R$ from $10^{-6}$~s$^{-1}$ to $10^{-3}$~s$^{-1}$ leads to
the decrease of the dimensionless coefficient $\tilde{\gamma_R}$
from $10^{-6}$ to $10^{-9}$. Therefore, in the dimensionless
units, the new magnon diffusion related term in the LL equation
has the coefficient which is large compared to the typically used
relaxation term. But, because this term contains derivatives of
the higher (second) order than other terms, its effect becomes
important only in the case of the strongly non-uniform states. For
example, for Walker's modes in the sample with the size of order
of 1~mm, the presence of the second derivative in the first term
in the left part of Eq.~(\ref{eq19}) decreases this term,
presented in the dimensionless units, by 13 orders of magnitude
and it becomes negligible.

Eq.~(\ref{eq19}) describes domains in the  $ZOX$ plane and the
magnetization oriented along the $z$ axis  and depending on $y$
($M\equiv M_z(y)$). Let us consider another orientation of the
domain plane. Using the free energy of  Eq.~(\ref{eq3}), we may
describe the domain in the $X0Y$ plane  with the magnetization
depending on $z$ ($M\equiv M_z(z)$) with an equation, which may be
obtained from Eqs.~(\ref{eq14},\ref{eq15},\ref{eq16}) by the
transformations $ M_z(y)\longrightarrow M_z(z), H\longrightarrow H-4\pi
M_z(z)$. The new term $4\pi M_z(z)$ may be combined with the term
$aM_z$ from the free energy. As the condition $a \gg 2\pi$ holds,
the equation for the magnetization for the alternative  $XOY$ orientation of the
domain plane will assume the form of Eq.~(\ref{eq18}).
Naturally, the resulting solutions will be similar. Therefore, we
shall study only Eq.~(\ref{eq18}) (Eq.~(\ref{eq19}) in dimensionless units).

\section{Investigation of stability of uniform magnetization}

At the uniform steady pumping, Eq.~(\ref{eq19}) has a uniform
steady state solution which determines the stationary
magnetization $M_0$ that satisfies the following equation
  \begin{equation}
\label{eq16a}
 M_0^3-M_0(1+q/\tilde{\gamma_R})-
 H=0.
 \end{equation}

 The solution of this equation in the absence of the pumping
 ($q=0$) determines the equilibrium magnetization $M_{eq}$.

\begin{figure}\centerline{\includegraphics[width=8.6cm]{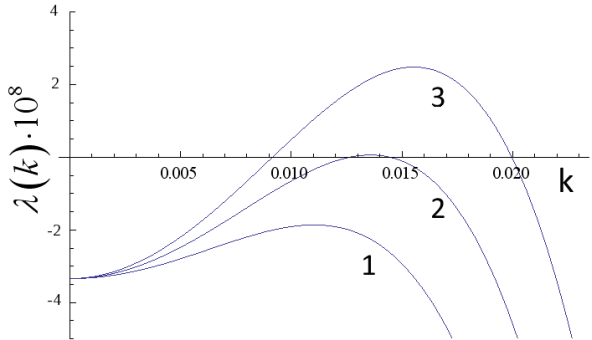}}\caption{ Dependence of the damping decrement $\lambda(k)$
on the wave number of the uniform solution fluctuation $k$  at the
relaxation rate $\tilde{\gamma}_R = 5\cdot 10^{-8}$, magnetic
field $H=0.001$ for different values of the pumping  parameter
$q$: 1) $3.3424\cdot 10^{-8}$, 2) $3.34261\cdot 10^{-8}$, 3)
$3.3428\cdot 10^{-8}$. The values are given in the dimensionless
units in this figure and in the subsequent figures.
 }\label{fig:fig2}\end{figure}

In order to investigate the stability of the uniform solution, let
us put $M=M_0+\delta M\exp(\lambda(k)t+iky)$, where $\delta
M \ll M_0$. The decrement of the damping obtained from
Eq.~(\ref{eq19}) is equal to
\begin{equation}
\label{eq17a}
 \lambda(k)=(1-3M^2_0-k^2)(k^2+\tilde{\gamma_R})-q.
 \end{equation}
Eq.~(\ref{eq17a}) implies, that the decrement $\lambda(k)$   is
negative in the absence of the pumping for every value of $k$,
meaning that the uniform state is stable. The dependence of the
damping decrement on $k$  is presented in Fig.~\ref{fig:fig2} for
the different values of the pumping $q$.  As seen from
Fig.~\ref{fig:fig2}, the damping decrement $\lambda(k)$  becomes
positive with increasing the pumping at a certain wave number
$k=k_c$.
At $q<3.3424\cdot 10^{-8}$  the uniform state is stable. At the
increased pumping $q>3.34261\cdot 10^{-8}$ the uniform solution
becomes unstable with respect to the creation of the periodical
magnon density variation with the wave number $k_c$. The
instability occurs in the region of the spinodal decomposition,
where the second derivative of the free energy with respect to the
order parameter ($M_z$ in the considered case) changes sign
($\partial^2 F/\partial M^2_z=0$).  But the numerical value of the
drop of the magnetization $M_z$ caused by the pumping needed for
reaching the instability is very large. For the example considered
in Fig.~\ref{fig:fig2},
 the magnetization $M_z$  changes from 1.0005 (in the
dimensionless units) at $q=0$ to 0.577244 at $q=q_s$. Such
a decrease of the magnetization requires an extremely strong
pumping, that is likely to change the temperature state of the
crystal. We shall not consider the region in the vicinity to the
spinodal decomposition and such strong pumping.

So, at $\lambda(k)<q_s$  the uniform state is stable even in the
presence of the pumping and for the finite lifetime of the
non-equilibrium state. But non-uniform stable states may coexist
with the uniform state even at $\lambda(k)<q_s$  in a certain
range  of the pumping intensity. These states arise when the
parameters of the system   belong to the region
 between the binodal and the spinodal. For a magnetic system,
the magnetization in this region is restricted by the conditions
imposed on $\partial F/\partial M=0$ and $\partial^2 F/\partial
M^2=0$ (this region is slightly affected by the pumping and the
finite value of the particle's lifetime). The condition $\partial
F/\partial M=-H_{eff}=0$ is realized in the equilibrium state and
determines the equilibrium magnetization. The region between
$\partial F/\partial M=0$ and $\partial^2 F/\partial M^2=0$ arises
at the magnetization smaller than the equilibrium value, that can
be caused by the pumping, which creates the magnons and decreases
the magnetization. The total number of magnons in this region
exceeds the equilibrium value. The system is supersaturated with
magnons. The gas of the particles in the state between the binodal
and the spinodal  may remain in the uniform supersaturated state
or could transfer due to fluctuations to the state with nuclei of
the new condensed phase. Subsequently, these nuclei grow with
time. In the case of stable particles, the growth of the condensed
phase nuclei slows down with time because  their number in the
matrix is limited. To the contrary, in a system of constantly
created particles with the finite lifetime,  the spatially
localized regions of the new phase  may be stabilized by the
interplay of the steady generation and decay. The stationary
localized islands of the condensed phase of indirect excitons
created by light in double quantum well heterostructures were
studied in \cite{Sug5, Dm} in the  parameter ranges at which the
uniform and the non-uniform solutions are stable simultaneously.
The localized solutions are called either the autosolitons (static
solitons) according to the classification by the paper\cite{Kern}
or the breathers according to \cite{Flach}. In the next section,
we shall use the Eq.~(\ref{eq19}) to study the non-uniform
non-equilibrium stationary states in the magnetic sample at the
steady pumping.

\section{Formation of a single domain}
We shall consider the steady state solutions of the
Eq.~(\ref{eq19}) for the magnetization. The non-uniform states
studied in the paper are domains with the magnetic moment oriented
opposite to the magnetization of the matrix. As have been
mentioned, we shall consider the solutions in the region on the
magnetization diagram where the uniform solution is stable, but
the non-uniform solutions appear as well due to the magnon created
by pumping. In the search of the non-homogeneous solutions, we
follow the procedure  applied in \cite{Sug5, Dm}. Two approaches
are used. In the first approach, we solve Eq.~(\ref{eq19}) for a
certain value of the pumping choosing the initial magnetization in
the form of the non-homogeneous function $M_z,(y,t)_{t=0}=
\emph{M}_{in}(y)$ depending on some parameters, the value of $q$
is given. In the second method, the external pumping is presented
in the form of $q\rightarrow q+Q_i(y,t)$, where $Q_i(y,t)$ is a
function containing some parameters and tending to zero at
$t\rightarrow\infty$, the initial magnetization being uniform.
Thus, the solution of the evolution  equation converges with time
to the solution of Eq.~(\ref{eq19}) with the given value of $q$.
Varying the functions $\emph{M}_{in}(y)$ and $Q_i(y,t)$, different
non-uniform stationary solutions for the magnetization may be
obtained. These solutions give $M(y,t)\rightarrow M_{eq}$ at
$q\rightarrow 0$. The solutions are stable because they do not
change at $t\rightarrow\infty$.  They do not change with the
variation  of the functions of $\emph{M}_{in}(y)$ and $Q_i(y,t)$
in some limited region of parameters of the functions. In other
words, for every solution there is a region of parameters of the
functions $\emph{M}_{in}(y)$ and $Q_i(y,t)$ in which the solution
is the same. This region determines "the attraction basin" for the
given solution.

In the absence of pumping ($q=0$), the solution of
Eq.~(\ref{eq19}) is uniform. Non-uniform solutions are possible at $q \neq 0$.
Firstly, let us find one such non-uniform
solution  localized in the center of the sample. It
may be obtained choosing the initial magnetization in the form
  \begin{equation}
\label{eq18a}
 \emph{M}_{in}(y)=\emph{M}_{in0}\exp[-(y-L/2)^2/s^2],
 \end{equation}
where  $L$  is the length of the sample, $\emph{M}_{i0}$ and $s$
are parameters.

\begin{figure}\centerline{\includegraphics[width=8.6cm]{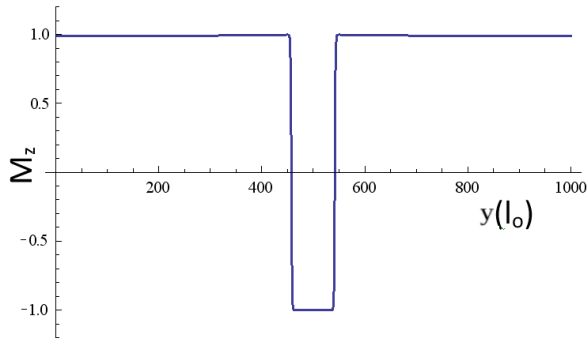}}\caption{
Spatial dependence of the magnetization at the pumping parameter
$q=0.000003$, the magnetic field $H=0.001$, the relaxation rate
$\tilde{\gamma_R}=0.0001$.
 }
\label{fig:fig3}\end{figure}

    If the function given by Eq.~(\ref{eq18a}) with certain values of parameters belongs to the attraction
basin of some solution of Eq.~(\ref{eq19}),  it converges to this
solution with time. We consider such a solution in the time limit
$t\rightarrow \infty$  as one of the desired solutions. The
example of the magnetization variation obtained in such way is
presented in Fig.~\ref{fig:fig3}. In the center of the sample, the
region is seen where the orientation of the magnetic moment is
opposite to the orientation of the magnetization of the remaining
part of the sample. The change of the parameters $\emph{M}_{in0}$
and $s$ of the trial function Eq.~(\ref{eq18a}) within  the some
limits gives the same solution of Eq.~(\ref{eq19}), presented in
Fig.~\ref{fig:fig3}. This confirms that the applied method for the
solution Eq.~(\ref{eq19}) is correct. Fig~\ref{fig:fig3} is the
manifestation of the scenario shown in Fig.~\ref{fig:fig1}. The
inverted spins (magnons) created by the pumping are clustered into
a domain.The applied method allows determination of the possible
states in the system. The realization of certain state depends on
the boundary initial condition and on fluctuations. The
calculations showed that
 the time of an establishment of the steady  size of
 the inclusions is very long (much longer than magnon
 lifetime). This fact should be taken into account performing an
 investigation of the magnon distribution at a pulse excitation.

The stationary state of the domain may exist if there is an
additional inflow of magnons from outside. There are minimal
values of the domain thickness and the sample thickness $L$ at
which the domain could develop. The  domain thickness  grows if
the value of $L$ rises, since the region from which the domain
harvests magnons increases. But, if $L$ becomes  greater than the
diffusion length, the size of the single domain reaches the limit
for the fixed value of the pumping rate. This occurs at
$L>(1/\gamma_R)^{1/2}$. In this limit, the results would not
depend on the boundary conditions. We studied the magnetization
for two types of the boundary conditions: for the periodical
conditions and for the fixed magnetization at the boundary.

\begin{figure}\centerline{\includegraphics[width=8.6cm]{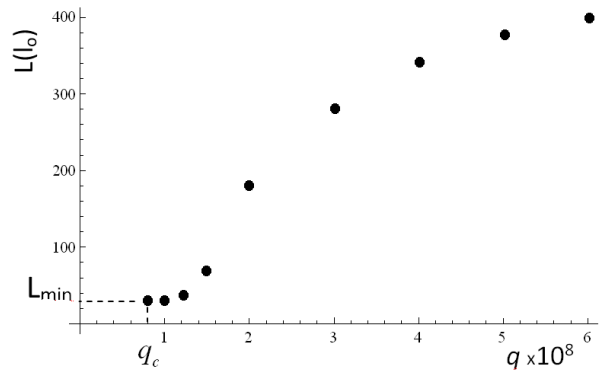}}\caption{
     Dependence of the thickness of the domain $L$ on the parameter of the pumping $q$. $q_c$ is the  pumping threshold of the domain
     creation, $L_{min}$ is the minimal thickness of the domain. The parameters of the system are: the relaxation rate
     $\tilde{\gamma_R}=10^{-6}$, the magnetic field $H=0.01.$}
\label{fig:fig4}
\end{figure}

The domain size grows with increasing the pumping.
Fig.~\ref{fig:fig4} presents the domain thickness as a function of
the pumping rate. It is seen from Fig.~\ref{fig:fig4}, that there
is a threshold of the pumping rate for the domain creation.
The pumping in Fig.~\ref{fig:fig4} leads to the decrease of the
magnetization. The pumping is such, that the relative change of
the uniform magnetization is equal to $5\cdot 10^{-3}$  and
$3\cdot 10^{-2}$   at $q=10^{-8}$  and $q=6\cdot 10^{-8}$,
correspondingly. As seen from Fig.~\ref{fig:fig4}, the threshold
$q_c$ for pumping is $0.8\cdot 10^{-8}$ for the given value of the  magnetic
field. Only the uniform solution exists at $q<q_c$. The threshold
grows with increasing the magnetic field. In the dimensional units,
at $T_c/(T_c-T)=10$, $q=5\cdot 10^{-8}$, $D=100$~cm$^2$/s, the
thickness of the domain reaches 1.2~m$\mu$. The size of the domains
increases with decreasing the parameter $\tilde{\gamma_R}$, {\it
id est}, with increasing the magnon lifetime.

If the decrease of the magnetization is such that the   state of
the system is  between the binodal and the spinodal, the nuclei of
a new phase (for example, a domain) are created due to
fluctuations. Only those nuclei survive which overcome a barrier.
Therefore, in order to describe this process of the domain
formation mathematically, we solved the main Eq.~(\ref{eq18a})
with the given initial trial non-homogeneous magnetization. If the
width of the initial non-homogeneous magnetization increases (the
value of $s$ in Eq.~(\ref{eq18a}) decreases), the solution with
two parallel domain arises. The two domains move apart. The
repulsion may be explained in the following way. Since  the
magnons in domains relax, the domains exist due to the magnon
inflow from outside. The region between the domains is common for
both domains and, as long as the number of the magnons is
restricted,
 their density is not sufficient to support the stationary state
 of two close domains. As a result, the domains
 tend to be situated on a certain distance from one another.

\section{Superlattice of domains}

\begin{figure}\centerline{\includegraphics[width=8.6cm]{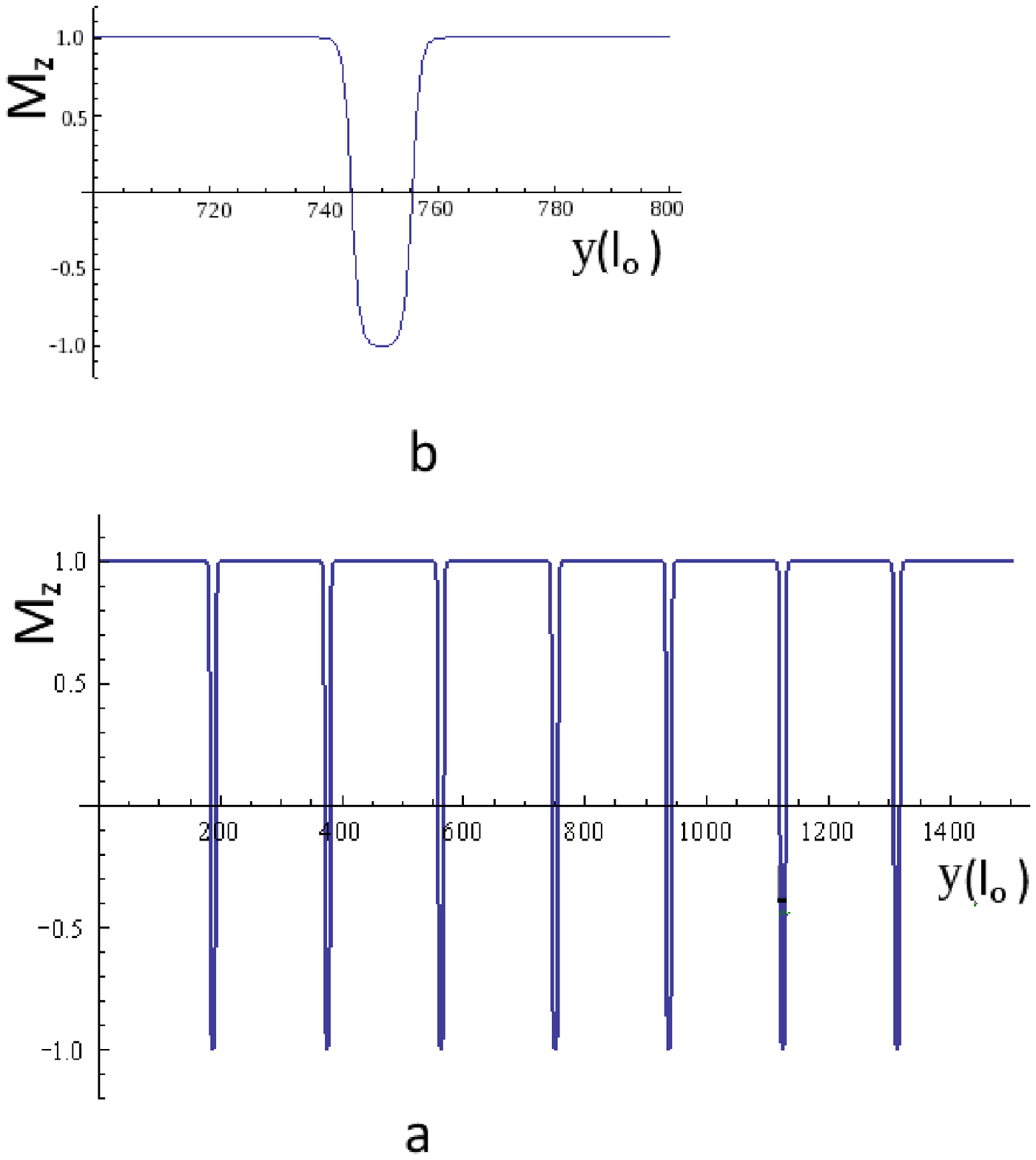}}\caption{Spatial dependence of the magnetization at
the pumping parameter $q=8\cdot10^{-9}$, the magnetic field
$H=0.01$, the relaxation rate $\tilde{\gamma_R}=10^{-5}$  (a). A
single domain is given separately in ($b$).}\label{fig:fig5}
 \end{figure}

\begin{figure}\centerline{\includegraphics[width=8.6cm]{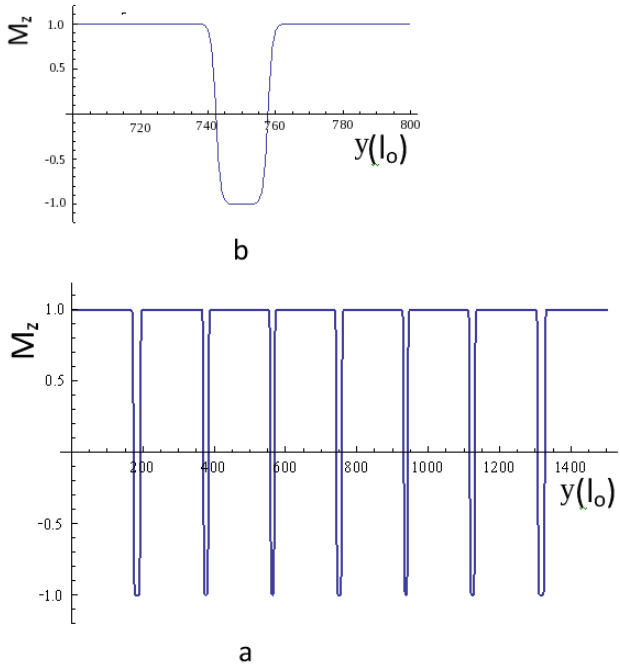}}\caption{
Spatial dependence of the magnetization at the pumping parameter
$q=1.5\cdot 10^{-8}$, the magnetic field $H=0.01$, the relaxation
rate $\tilde{\gamma_R}=10^{-5}$.($a$). A single domain is given
separately in ($b$).} \label{fig:fig6}
 \end{figure}

Solutions of Eq.~(\ref{eq19}) with the periodical
variation of the magnetization are also possible. Some of them are presented in
Figs.~\ref{fig:fig5} and \ref{fig:fig6}. In Figs.~\ref{fig:fig5}b
and \ref{fig:fig6}b, the enlarged regions of one of the domains
are given separately.

\begin{figure}\centerline{\includegraphics[width=8.6cm]{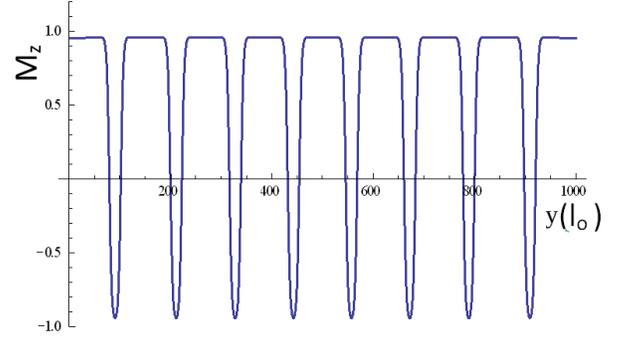}}\caption{
Periodical distribution of the magnetization at pumping parameter
$g=0.018$, the magnetic field
 $H=0.005$, the relaxation rate $\tilde{\gamma_R}=0.2$.} \label{fig:fig7}
 \end{figure}

Comparing Fig.~\ref{fig:fig5}, Fig.~\ref{fig:fig6} and
Fig.~\ref{fig:fig7}, one can see that with increasing the pumping
rate the domains widen.

 The appearance of a periodical structure of the magnetization  was observed in \cite{Dem4}  in YIG films at the external pumping.

Some comments about the effect of fluctuations are possible. The
problem of the fluctuations of the density of unstable particles
were investigated  in the papers\cite{Sug3, Ogawa, Ish} by solving
the Fokker-Planck functional equation for the free energy in the
Landau-Ginzburg form. Both the intrinsic fluctuations and the
fluctuations of the pumping were taken into account.  The studies
shown the appearance of a second  maximum in the one-point
distribution function if the  pumping rate exceeded a certain
value. This maximum corresponded to the development of the second
phase. The separation of two phases may occur if the particle
lifetime is greater than a certain threshold value. The two-point
correlation function was obtained.  The  Fourier transform of
two-point correlation function calculated in \cite{Sug3}  has a
sharp maximum at some value of the wave vector that corresponds to
the oscillations of the correlation function as a function of
coordinates. So, periodical structures appear in the system. The
problem has solved in \cite{Sug3}  differs from the problem in the
current work by  the presentation of the lifetime of  particles:
the lifetime in \cite{Sug3} is constant, it is the function of the
spatial coordinate (the second term in right hand side of
Eq.~(\ref{eq19}).  But  the first term in the right hand side of
Eq.~(\ref{eq19})  plays the main role in the formation of
non-uniform structure.  Using these results, we may argue that the
fluctuations do not destroy the periodical structures that arise
in the considered ferromagnetic system, if the magnon lifetime is
much greater than the times of magnon-magnon and magnon-phonon
collisions and  the quasi-equilibrium state  is formed in the
system.

 \section{Discussion and conclusion}

We have shown that, in the system with the high density of
quasi-equilibrium magnons, non-uniform structures  may arise
similar to the formation of inclusions of the condensed phase in a
supersaturated gas or to the separation of the new phase in the
crystal supersaturated with impurities. Performing the
calculation, we used the Landau presentation of the free energy of
magnons, which is valid in the vicinity to the phase transition
temperature. The work explored the simplest model, namely, we
considered the appearance of simple defects in the form of
domains. But, in our opinion, some of the obtained results give
insight into the general behavior in more complicated cases.
Besides the considered simplest one-dimensional structures, the
system may develop other types of non-uniformities having
different shapes and being restricted in all directions (disks,
balls, ellipsoids, and  so on). What new features  may be expected
when more complex regions of the new phase are formed? Let us make
some qualitative analysis of the possible manifestations of the
development of non-uniform structures. The following list names
the most important of them.
\begin{enumerate}
\item The appearance of the regions of the non-uniform magnetization
causes additional scattering of the electromagnetic waves. The
intensity of the light scattering by clusters  of magnons at some
frequencies is greater than by the same number of the individually
independent particles.

\item Since the
non-uniform inclusions are small,  the magnon levels in them will
be quantized. The lowest states vary slowly in space  and manifest
themselves intensively in the scattering and absorption of the
electromagnetic waves. The levels with small quantum numbers  will
display themselves most strongly. This may cause an increase in
the lowest part of the scattering spectra similar to the effect
observed at the Bose-Einstein condensation. It was shown that
stationary state becomes established if the time of the pumping
pulse duration is much longer that the magnon lifetime. The sizes
of  inclusions will growth with  the increasing duration of the
pulse. It will cause to the step-down of the quantized levels of
magnons in the inclusions and to the shift of the scattering
spectra of electromagnetic waves to the lower frequencies.

\item The spin orientation in an inclusion of the condensed phase is
  determined by the strong exchange
interaction. In order to change the orientation of its spin to the
orientation of  the matrix spins, a magnon should jump out of the
inclusion, where it is bound by the strong exchange interaction.
There is another mechanism of the spin relaxation in the inclusion.
Firstly, magnons (light arrows in dark regions of
Fig.~\ref{fig:fig1}) are exited in the domain and then leave the
inclusion. This is a two-stage process and therefore it has small
probability.  As the result the inclusions should live long.

\item The stationary distributions of the number of inclusions and their sizes arise under stationary pumping.
 In this case the number of  magnons captured by the inclusions coincides   with the number of magnons, which leaves the inclusions
 and decay in inclusions. If the magnon pumping goes off, the sample cooling occurs more quickly than the decrease
 the magnon number. As the result, a some excess of magnons will exist some time in comparison with equilibrium value.
 Because the number of magnons leaving inclusions will decrease due to the temperature decrease, the number of magnons
 in the inclusion will rise. Later the excess of magnons drops and inclusions disappear. So, the peak of magnon number
 in inclusions may arise after switching off the pumping.  Such peak was observed during investigation
  of Bose-Einstein condensation of magnons \cite{Serga}. 

\item The regions of the condensed phase in systems supersaturated
  with magnons may arrange themselves into structures.  These
  structures would form a periodical distribution of domains with
  parallel planes. The normal to the domain plane may be oriented either along the external magnetic field or perpendicular to the field.
 Such picture was observed  in Ref.\cite{Dem4}.
\end{enumerate}

 All these effects, which may be expected due to the inclusions of the
 new phase,
 were observed in experiments
 \cite{Dem1,Dem2,Dem3,Dem4}, and explained as the manifestation of the Bose-Einstein condensation.
 We cannot object to that attribution because our study is
 carried out in  somewhat different conditions. The majority of experiments \cite{Dem1,Dem2,Dem3} were
 carried out
 using the magnon excitation by pulses. Our calculations studied the steady states.
   Yet we would like to
 underline
  that, besides the Bose-Einstein condensation,  an another scenario should be considered
  when studying the many-magnon systems.  A search  for the inclusions of
  the new phase and evaluation of their role in the physical processes should be performed.

Thus, the presented paper studied a magnetic sample  with a large
magnon concentration created by the external pumping. Due to the
long lifetimes, magnons are in the quasi-equilibrium state.  It is
shown that besides the Bose condensation, the formation of the
inclusions of new condensed magnon phase may occur similarly to
the development of the condensed phase in a supersaturated gas.
The exchange interaction promotes the combining the separate
magnons into inclusions.
 The orientation of magnetic
moments in each inclusion is opposite to the magnetic moment of
the magnetic matrix. Therefore, the inclusions may exist only in
condition of the external pumping. Summarizing, inclusions are dissipative
structures that arise as the result of self-organization
processes  in non-equilibrium conditions\cite{N}. The
presented study considered the inclusions  of the condensed phase
shaped as individual domains or periodical supperlattices of
domains.

 The possibility of the formation
 of the inclusions of the new phase  should be taken into account side by
side with the process of Bose-Einstein condensation when analyzing
experiments.

The author is grateful to B.A. Ivanov, V.M. Loktev,
S.M.~Ryabchenko, G.A.~Melkov, and I.Yu.~Goliney for fruitful
discussions.

\end{document}